\documentclass[11pt]{article}
\usepackage{hyperref}
\pdfoutput=1

\begin{document}
\title{\textbf{Exploding Taylor Cones}}
\author{John C. Creasey, Brad S. Hamlin, \& William D. Ristenpart \\
\\\vspace{6pt} Dept. Chemical Engineering \& Materials Science, \\ University of California at Davis, Davis, CA 95616, USA}
\date{October 15, 2010}
\maketitle

Application of a sufficiently strong electric field to an aqueous solution induces a phenomenon known as `electrohydraulic discharge' [1]. The
electric field causes the water to break down, generating either a corona (at lower field strengths) or a pulsed arc (at higher field
strengths).  The discharge typically results in a complex combination of physical processes (e.g., cavitation and light emission) and chemical
reactions (e.g., generation of free radicals and nonthermal plasmas).   The combination of physical and chemical processes tends to destroy any
organic molecules present, and accordingly electrohydraulic discharges are currently being investigated as a potentially inexpensive and
environmentally friendly means for purifying drinking water and removing contaminants from wastewater [1].

Two types of electrode configurations have been the main focus of research to date: (i) a `one-phase' system with both electrodes immersed in
water, or (ii) a `two-phase' system with one electrode in air and the other submerged in water. In this \href{http://arxiv.org}{fluid dynamics
video}, we demonstrate the striking consequences of triggering an electrohydaulic discharge in a two-phase system comprised of water and a
viscous, insulating oil. An air/water interface typically remains stationary until the discharge occurs; in contrast, the oil/water interface
deforms into a conical shape (i.e., a Taylor cone [2]) stretching from the water phase toward the oil-immersed electrode. The behavior after the
cone contacts the electrode depends sensitively on the properties of the water and oil, and we demonstrate that, under appropriate conditions,
destructive explosions occur.

The experimental apparatus is similar to that used by Ristenpart \emph{et al}. [3], in which the bottom half of a plastic container (1x1x5 cm)
is filled with water and the top half filled with oil. Metal wires are inserted into each liquid (at top and bottom) to serve as electrodes. A
standard high voltage power supply (Trek 610E) provides an electric potential difference on the order of 3 kV over approximately 1 cm. No
special circuitry is used to generate a rapid rise time in the field strength; because of the insulating oil, the current density is effectively
zero until the Taylor cone contacts the electrode, providing ample time for the applied potential to reach its specified value.  The movies were
recorded using a Phantom v7.3 camera capturing at 14,000 frames per second (fps).  The entire apparatus (excluding camera) was enclosed in a
transparent `blast shield' composed of 1-inch thick plexiglass.

The fluid dynamics video contains several segments that illustrate the observed behavior.  The first segment demonstrates the growth of a Taylor
cone at relatively high magnification, played back at 20 fps.  The water contains 1 M KCl, the oil is 500 cSt polydimethyxlsiloxane (PDMS), and
the applied potential is 3 kV.  Note that at the late stages, the `cone' tip stretches toward the electrode and makes contact.

The behavior after contact is demonstrated in the next series of movie segments, which illustrate four separate experiments: three with various
oil/water combinations, and one with air/water to provide contrast.  The experimental details are as follows:
\begin{description}
  \item[(1) Low Field, High Viscosity, High Salt:] The applied potential is 3 kV, the oil viscosity is 500 cSt, and the salt concentration in the water is 1
  M KCl.  Upon contact, an explosion occurs with a shockwave propagating radially outward from the electrode tip. The shockwave ultimately causes
  the plastic container to fail along the seams at each corner. The field of view is 2.5 x 5.0 cm, and the original capture rate was 14,000 fps played back here at 10 fps.

  \item[(2) High Field, Low Viscosity, High Salt:] The applied potential is 4.5 kV, the oil viscosity is 10 cSt, and the salt concentration is 1
  M KCl.  Under these conditions a qualitatively more powerful explosion occurs compared to experiment 1.  The force of the explosion is more localized, effectively
  shattering the plastic container in the vicinity the electrode tips.  Note that a few `glowing debris' fly through the field of view;
  the nature of these objects is unclear.The field of view is 2.5 x 5.0 cm, and the original capture rate was 14,000 fps played back here at 20 fps.

  \item[(3) Low Field, High Viscosity, Low Salt:] The applied potential is 3 kV, the oil viscosity is 500 cSt, and the salt concentration is
  $10^{-3}$ M KCl.  Upon contact, a pronounced `arc' develops in the tip of the Taylor cone, which subsequently explodes.  At this lower salt concentration,
  however, the resulting explosion is much weaker: the plastic container does not fracture or shatter. The field of view is 1.3 x 2.7 cm, and the original capture
  rate was 14,000 fps played back here at 20 fps.

  \item[(4) High Field, Inviscid (air), High Salt:] The applied potential is 3 kV, the insulating fluid on top is air, and the salt concentration is
  1 M KCl.  In contrast to the oil/water interface, at these field strengths the electric driving force is insufficient to deform the
  air/water interface into a Taylor cone.  Instead, the interface remains flat and eventually dielectric breakdown of the air occurs, ultimately
  causing the water to `splash' up and around the electrode.  Multiple dielectric breakdown events (`sparks') continue to occur chaotically as the water
  droplets `short out' the system intermittently, but ultimately the plastic container is undamaged. The field of view is 1.3 x 2.7 cm, and the original
  capture rate was 14,000 fps played back here at 75 fps.

\end{description}
The above observations raise several interesting questions about shockwave propagation in viscous multiphase fluids and about the influence of
ionic conductivity on arc generation within Taylor cones.  For anyone interested in reproducing these experiments, however, we offer the
following notes of caution:
\begin{enumerate}
  \item The explosion can cause jagged shards to fly across the room ($>20$ feet).  The use of a blast shield and eye protection is strongly recommended.
  \item The explosion itself causes a startlingly loud `cracking' noise.  Ear protection is recommended.
  \item Very high current densities ($>100$ mA) occur transiently during the explosion.  Standard precautions for handling high voltages and
  currents are imperative to prevent electrocution.
\end{enumerate}
Additional movies illustrating the phenomena are available upon request.  Please contact William Ristenpart at wdristenpart@ucdavis.edu for more
information.

\newpage

\noindent [1] Locke, B. R., Sato, M., Sunka, P., Hoffmann, M. R. \& Chang, J. S. ``Electrohydraulic discharge and nonthermal plasma for water
treatment,'' \emph{Industrial \& Engineering Chemistry Research} \textbf{45}, 882-905 (2006).

\vspace{0.5cm}

\noindent [2] de la Mora, J. F. ``The fluid dynamics of Taylor cones,'' \emph{Annual Review of Fluid Mechanics} \textbf{39}, 217-243 (2007).

\vspace{0.5cm}

\noindent [3] Ristenpart, W. D., Bird, J. C., Belmonte, A., Dollar, F. \& Stone, H. A. ``Non-coalescence of oppositely charged drops,''
\emph{Nature} \textbf{461}, 377-380 (2009).

\end{document}